\documentclass[a4paper]{mem}
\usepackage{natbib}
\usepackage{graphicx}
\usepackage{txfonts}
\usepackage[a4paper]{hyperref}
\idline{73}{23}

\begin{document}

   \title{The XMM Large Scale Structure (XMM-LSS) Survey:
	  X-ray analysis and first results }

   \author{A.M. Read \inst{1}
\thanks{on behalf of the LSS consortium } }

   \offprints{A.M. Read}
\mail{Dept. of Physics and Astronomy, Leicester University, 
      Leicester LE1 7RH, U.K. }

   \institute{Dept. of Physics and Astronomy, Leicester University, 
              Leicester LE1 7RH, U.K.}

   \abstract{ Thanks to its unprecedented sensitivity and large field
of view, XMM occupies a leading position as a survey instrument. The
XMM-LSS survey is a medium-deep large-area X-ray survey, and greatly
extends the cosmological tests attempted using earlier studies. While
a wealth of complementary multiwavelength studies are taking place, I
concentrate here on the XMM-EPIC data, describing aspects of the X-ray
pipeline and analysis, and selected first results, including the
deepest wide-angle X-ray image of the cosmos to date.

 } 

\authorrunning{A.M. Read} 

\titlerunning{The XMM Large Scale Structure (XMM-LSS) Survey} 

\maketitle

  \begin{figure*}
  \centering
  \resizebox{\hsize}{!}{\rotatebox[]{-90}{\includegraphics[bb=63 37 527 772,clip]{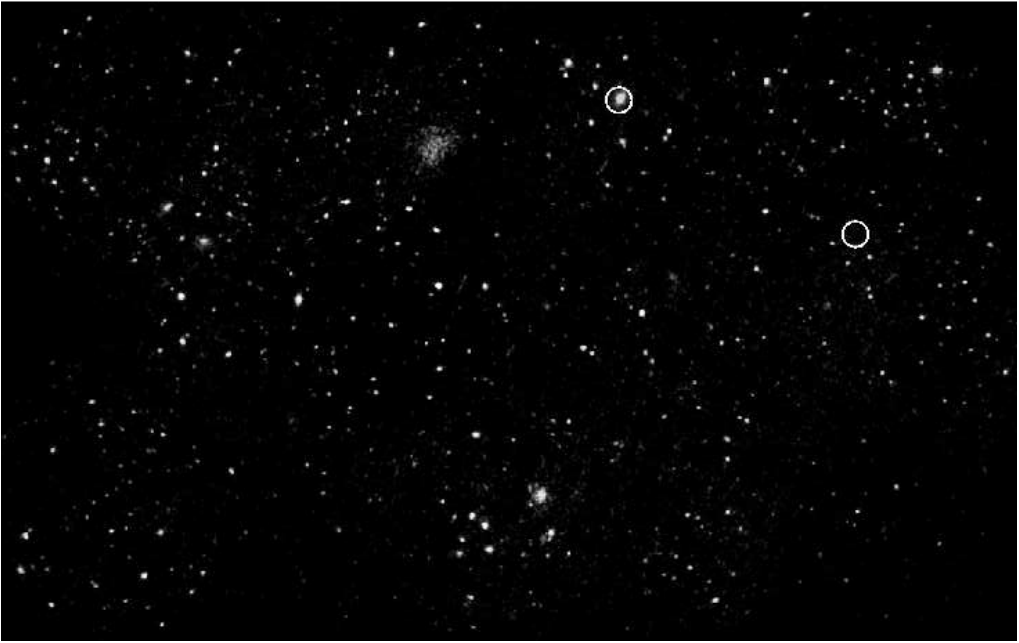}}}
   \vspace{-28mm}
  \caption{{\bf Deepest large-scale view of the X-ray sky.}
  1.6\,deg$^{2}$ mosaic of the first 15 10\,ks all-EPIC XMM-LSS fields. 
  Circles indicate sources found in the RASS (the brightest source is a star; HD14938).
  (3-colour image at http://vela.astro.ulg.ac.be/themes/spatial/xmm/LSS/First/FMos/). 
  }
  \label{Fig1}
  \end{figure*}

\section{Survey motivation and design}

An XMM-EPIC medium-deep large-area X-ray survey - the XMM Large Scale
Structure survey (XMM-LSS) \citep{LSS1} - has been designed to extend
earlier ROSAT cosmological tests \citep{bohr} to two redshift bins
between 0$<$$z$$<$1, while maintaining a high precision. The
evolutionary study of the cluster-cluster correlation function and of
the cluster number density has constrained the survey design: 
an equatorial mosaic of 10\,ks pointings, separated by 20', and
covering $8^{\circ}\times8^{\circ}$, giving a sensitivity of $\sim$$
3\times10^{-15}$\,erg cm$^{-2}$ s$^{-1}$ in the 0.5$-$2\, keV band,
and yielding $>$800 clusters and many more AGN with a space density of
$\sim$250\,deg$^{-2}$. The survey will be able, for the first time,
to map out the the evolution of the LSS of the universe out to
$z$$\sim$1 with both the galaxy cluster and the QSO populations.

Several other major surveys from the radio to the UV wavebands are
currently being undertaken over the LSS sky region, including the
CFHTLS weak lensing and the AMiBA SZ surveys, Magellan and VLT/FORS2
spectroscopic surveys, and CTIO, UKIRT, VLA, OCRA, SIRTF and GALEX
studies (plus others). Constraints of space allow only a discussion
here of the XMM-EPIC data. The wide scope of the LSS survey has
necessitated the assemblage of a large consortium, and the project is
presented in detail at \verb+http://vela.astro.ulg.ac.be/themes/+
\verb+spatial/xmm/LSS/index_e.html+.

\section{X-ray analysis and first results}

Fig.\,1 shows a mosaic of the first 15 XMM-LSS fields. The improvement
over the RASS is very striking, with a source density of
$\sim$300\,deg$^{-2}$ in the 0.5-2\,keV band. Supersoft and very hard
sources are seen, as well as sources covering a wide range in extent,
all indicating the scientific potential of the survey. The large,
extended feature to the north (XMMUJ022540-031111) is associated with
a $z$$\sim$0.3 merging cluster, the single X-ray peak (hotter than the
surrounding cluster emission) located {\em between} the two galaxy
subclumps.

The XMM-LSS pipeline is based on a 3-stage
filtering/detection/measurement process. Photon images are
wavelet-filtered (in counts, to preserve Poisson statistics) for each
EPIC instrument separately. The three exposure-corrected images are
then summed for a first Sextractor detection step, the output being
fed into a final SAS-emldetect step. Tests involving dedicated
detailed simulations, incorporating all instrumental effects,
background and both point and extended sources, show that most, if not
all ($T$$>$4\,keV) clusters within $z$$\sim$1, and the brightest clusters
up to $z$$\sim$2 (if they exist) should be detected in a 10\,ks
exposure. Further, a very small rate ($<$1 per pointing) of spurious
detections is expected.

Initial results, described in detail in \citet{LSS1}, \citet{LSS2} and
also Willis et al. \& Andreon et al. (in preparation), indicate that
we are detecting, as expected (for a $\Lambda$CDM cosmology),
$\sim$15+ extended source per square degree, with counterparts in the
optical and in the NIR, and we are observing clusters beyond $z$$>$1. We
see that a 10\,ks XMM exposure and 2$h$ on the VLT allows a $kT$ (to
$\pm1$\,keV) and $\sigma$ measurement out to $z$$\sim$0.8. Many
intermediate mass systems are being uncovered, as well as relaxed,
merging and collapsing clusters, and many AGN.

\begin{acknowledgements}
  AMR would like to thank the entire LSS consortium
\end{acknowledgements}

\bibliographystyle{aa}

\begin{thebibliography}{}

\bibitem[B\"{o}hringer et al.\,(2001)]{bohr}
B\"{o}hringer H., et al., 2001, A\&A 369, 826
\bibitem[Pierre et al.\,(2003)]{LSS1}
Pierre M., et al., 2003, astro-ph 0305191, A\&A accepted
\bibitem[Valtchanov et al.\,(2003)]{LSS2}
Valtchanov I., et al., 2003, astro-ph 0305192, A\&A accepted

\end{thebibliography}

\end{document}